\documentclass[conference]{IEEEtran}
\IEEEoverridecommandlockouts
\usepackage{cite}
\usepackage{amsmath,amssymb,amsfonts}
\usepackage{graphicx}
\usepackage{textcomp}
\usepackage{xcolor}
\def\BibTeX{{\rm B\kern-.05em{\sc i\kern-.025em b}\kern-.08em
    T\kern-.1667em\lower.7ex\hbox{E}\kern-.125emX}}
\begin{document}

 \title{LSU factorization
}
 
\author{\IEEEauthorblockN{  Gennadi Malaschonok${}^{*}$ \thanks {${}^{*}$This is preprint of the paper: G. Malaschonok, "LSU Factorization," 2023 International Conference on Computational Science and Computational Intelligence (CSCI), Las Vegas, NV, USA, 2023, pp. 472-478, doi: 10.1109/CSCI62032.2023.00083.} \thanks{ ${}^{**}$This work was supported by a grant from the Simons Foundation, ID:1030285. The conference registration fee was paid by the Ukrainian company Genesis.}  
}
\IEEEauthorblockA{\textit{Faculty of Computer Science} \\
\textit{National University of Kyiv-Mohyla Academy}\\
 Kyiv, Ukraine\\
malaschonok@ukma.edu.ua} 
}

\maketitle

\begin{abstract}
The matrix LU factorization algorithm is a fundamental algorithm in linear algebra. We propose a generalization of the LU and LEU algorithms to accommodate the case of a commutative domain and its field of quotients. This algorithm decomposes any matrix A into a product of three matrices A=LSU, where each element of the triangular matrices L and U is a minor of matrix A. The number of non-zero elements in matrix S is equal to rank(A), and each of them is the inverse of the product of a specific pair of matrix A minors. The algorithm's complexity is equivalent to that of matrix multiplication.
\end{abstract}


  \newtheorem{theorem}{Theorem}  
   \newtheorem{definition}{Definition}  
   \newtheorem{property}{Property}

\section{Introduction}

The LU factorization represents matrix $A$ as a product, $A = LU$, where L is a lower triangular matrix, and U is an upper triangular matrix. It serves as a foundational algorithm in linear algebra program libraries \cite {1}, and various algorithms, including solving linear systems, matrix inversion, and rank calculation.

With supercomputers, handling larger matrices in applied problem-solving has become possible, but limitations of matrix algorithms have also become evident. To effectively use big data, issues like rounding errors, loss of precision, poor parallelism, and high computational complexity must be addressed (see, for example, \cite {2}).

As matrix sizes increase, computational errors also grow. Today's calculation limitations are not due to insufficient computing power of supercomputers, but rather the accumulation of computational errors. As such, the LEU algorithm (\cite {3}), a block recursive algorithm with matrix multiplication complexity, is efficient for sparse matrices and its coarse-grained algorithm parallelizes well. However, accumulating computational errors for matrices over classical fields ${\mathbb R}$ and ${\mathbb C}$ remains a challenge. 

We consider matrices over commutative domains and their quotient fields and propose a generalization of the LEU algorithm, which we call the LSU algorithm. Simultaneously with the calculation of the factors of the original matrix $A$, the factors of the matrix $P$ inverse to $A$ are also computed. If the original matrix $A$ is not invertible, then $P$ is a pseudoinverse matrix for which two identities hold:
$ A=APA \hbox{   and   } P=PAP. $

In what follows, we propose a new factorization algorithm, 
demonstrate its existence and correctness, and estimate the total number of operations required. Additionally, we provide a detailed example of the LSU factorization.
Preliminary results in this direction were previously obtained 
in papers \cite{4} - \cite{7}. 
The LSU factorization algorithm serves as another step towards creating a comprehensive 
library of block-recursive linear algebra algorithms.

A.A. Karatsuba and W. Strassen achieved the initial results in this field, but the importance of recursive algorithms for supercomputer computations has only been recognized in recent decades, spurring a program to develop decentralized dynamic control technology for supercomputer processing (see \cite{2022a}, \cite{2018I}). Other examples of recursive algorithms in the commutative domain, such as computing inverse and adjoint matrices, and the kernels of a linear operator, can be found in \cite{2006} and \cite{2018I}.

\section{Initial statement of the problem}
 Let $R$ be a commutative domain, $F$ its field of quotients. Let $A\in R^{n\times n}$  be a matrix that has rank $r$, $ r\le n$. 
 Let $ det_{r}, det_{r-1},..,det_{1} $ be some sequence consisting of $r$ nested nondegenerate minors of matrix $A$, which have sizes $r,r-1,..,1$, respectively.
 We want to get matrices $L,U \in R^{n\times n}$ of rank $n$,( L is  lower triangular, U is upper triangular), matrix $S $, that has rank $r$,  with $r$ non-zero elements equal to $(\ det_{1})^{-1}$, $(det_{1} det_{2})^{-1}$, .., $(det_{r-1} det_r)^{-1}$ such that 
  $$ A=LSU. $$ 
To solve this problem, we will formulate a more general problem, but first we will give some necessary definitions.
 \section{ Preliminaries}
 \subsection{ Semigroup of  weighted permutations}
The set of all matrices from $F^{n\times n}$ that contain at most one nonzero element in each row and each column forms a weighted permutations semigroup ${\bf S}_{wp}$.

The diagram at Fig.1 shows the structure of this semigroup.

  \begin{figure}[tb]
\begin{center}
 
\includegraphics[scale=0.5]{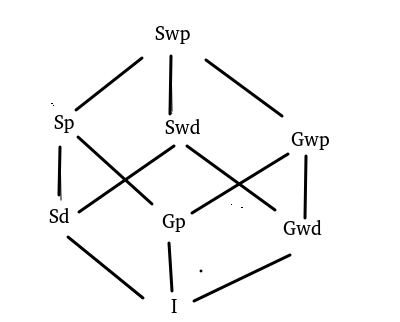}
  \caption{The structure of the  weighted permutations semigroup ${\bf S}_{wp}$}
 \label{graphinv}
 \end{center}
\end{figure}

In the center of the diagram we see the permutation group ${\bf G}_p$. Between ${\bf S}_{wp}$ and ${\bf G}_p$ there are two more subalgebras: ${\bf S}_p$ and ${\bf G}_{wp}$.
 
 If matrices from the permutation group ${\bf G}_p$ are supplemented with matrices that are obtained by replacing identity elements with arbitrary nonzero elements, then we get a group of weighted permutations ${\bf G}_{wp}$.
 
 If, on the contrary, to the matrices from the permutation group ${\bf G}_p$ we add the matrices that will be obtained after replacing some identity elements with zero ones, then we get the permutation semigroup ${\bf S}_p$.
 
The weighted permutation semigroup ${\bf S}_{wp}$ is obtained from the group ${\bf G}_{wp}$ by adding matrices, which are obtained after replacing some non-zero elements with zeros.

All diagonal matrices in the semigroup ${\bf S}_{wp}$ form a semigroup of weighted diagonal matrices
${\bf S}_{wd}$. Two other subalgebras ${\bf S}_d$ and ${\bf G}_{wd}$ are embedded in it.

The semigroup $ {\bf S}_d $ is formed by diagonal matrices that can have only 1 and 0 on the diagonal. The group ${\bf G}_{wd}$ is formed by those diagonal matrices from ${\bf S}_{wp }$ that do not have zero elements on the diagonal.

The identity group $ \bf I $ 
completes this construction.

 \subsection{Some mappings on semigroups}
  For matrices from the semigroup $ {\bf S}_{wp} $ we introduce two mappings:    {\bf unit} and {\bf extended}.

  A homomorphism of the multiplicative group $ F^* \to 1 $ (and semigroup $ R^* \to 1 $)  induces a homomorphism of the corresponding subalgebras:
  ${\bf S}_{wp} \to {\bf S}_{p}$, ${\bf S}_{wd} \to {\bf S}_{{d}}$, ${\bf G}_{wp} \to {\bf G}_{p}$, ${\bf G}_{wd} \to I$. 
All nonzero elements of the matrix are replaced by unit elements.  This corresponds to the communication lines in the diagram, which
 are directed down and to the left.
 \begin{definition} 
   The mapping of the matrix 
   $ S \in {\bf S}_{wp} (F) $ induced by the homomorphism 
   $F^* \to 1$:
 $$  S\in {\bf S}_{wp} \to  S^{\to 1} \in {\bf S}_{p}$$
 is called   {unit mapping}. 
  \end{definition}
The unit homomorphism of semigroups can be represented by the following commutative diagram:


$\hspace{40mm}    (A,B) {\longrightarrow} (  A^{\to 1}, B^{\to 1})$

 $\hspace{40mm}   \  \times \downarrow  \hspace{17mm}  \times \downarrow  $

 $\hspace{45mm}      C \ \ \ { \longrightarrow} \ \ \ \ \hspace{4mm}  C^{\to 1}$

\begin{definition}
The matrix mapping $S\in {\bf S}_{wp} \to   S^{Ext} \in {\bf G}_{wp} $ in which the block at the intersection of zero rows and zero columns is replaced by a unit block called {\bf extended mapping} and is indicated by a "Ext" upper index.
\end{definition} 
 On the diagram, the extended mapping corresponds to 4 arrows that are directed down and to the right:   ${\bf S}_{wp} \to {\bf G}_{wp}$, ${\bf S}_{wd} \to {\bf G}_{wd}$, ${\bf S}_{p} \to {\bf G}_{p}$, ${\bf S}_{d} \to I$.  
 As a result of such a transformation, a matrix of full rank is obtained.
 
 \begin{definition}
    The mapping of the matrix $S\in {\bf S}_{wp} \to  \bar  S =S^{Ext}-S \in {\bf S}_{p}$ 
    is called the {\bf  complementary mapping} and is denoted by a ``bar''.
 \end{definition}

\begin{property}
The special case of the complementary mapping, when the matrix $S$ belongs to the semigroup ${\bf S}_p$, is an involution on the semigroup ${\bf S}_p$. Involution is reversible: $\bar {\bar S}=S$. 
\end{property}%

Examples of  involution:  

$$\left[\begin{array}{ccc}0  & 0 & 0\\ 0  & 0 & 0\\ 1  & 0 & 0 \end{array}\right] 
\to  \left[\begin{array}{ccc}0  & 1 & 0\\ 0  & 0 & 1\\ 0  & 0 & 0  \end{array}\right]
.$$
\begin{property} 
$ \forall S \in {\bf S}_p: S+\bar S \in {\bf G}_p $.
\end{property}%
   To denote the Moore-Penrose generalized inverse matrix, we use the superscript plus.  It is easy to check the following property.
    \begin{property}
   For any matrix   $ S=(a_{i,j})\in {\bf S}_{wp}$ the Moore-Penrose generalized inverse matrix is equal   
   $$
   S^{+}=\{(b_{i,j}):b_{i,j}=0\hbox{ if }a_{j,i}=0, b_{i,j}=a_{j,i}^{-1} \hbox{ if }a_{j,i}\ne 0\}.
   $$   \end{property} 
 
   The matrix $S^{+}$ is obtained by transposing the matrix $S\ (S\in {\mathbb Z}_{wp})$ and replacing each non-zero element by the inverse element.

\subsection {Surrounding  minors}

  Hereinafter, we consider matrices over the commutative domain $ R $.
\begin{definition}
Let a matrix $ \mathcal M $ be given and let $ A $ be a square block located in the upper left corner of the matrix $ \mathcal M $. Any submatrix $ G$, 
$$ G= \left[\begin{array}{cc}A  & c \\ r & \omega \end{array}\right],  \eqno(1) $$
   that is obtained by adding to the block $ A $ some row $ r $ and some column $ c $ of matrix $ \mathcal M $ is called the submatrix that surrounds the block $ A $. Its determinant we call the surrounding minor.
 \end{definition}
 
 \begin{theorem}
 Let $ A $ be a square matrix, $ A^* $ the adjoint matrix for $ A $, $ det (A) \ne 0 $, $ G $ (1) be the surrounding  matrix for $A$, then 
 $$det(G)=  det(A) \omega - r A^* c.$$
 \end{theorem}
 %
 {\bf Proof.}
 This equality expresses the decomposition of the determinant $ G $ in row $ r $ and column $ c $.
 %
 $\square$
\begin{theorem} 
Let a matrix $ \mathcal M $ be divided into blocks
$$ {\mathcal M} =  \left[\begin {array} {cc} A_k & B \\ C & D \end{array} \right], $$ $ A_k $ is a square block of size $ k \times k $, its determinant $ det_k $ is non-zero and $ A_k^* $ is an adjoint matrix for $ A_k $, then each element of the matrix $A$,
  $$ A = (det_k) D - C A_k^*B, $$
is a surrounding minor for the $ A_k $ block.
 \end{theorem}
 {\bf Proof.} 
 To prove it, we need to apply Theorem 1 to each element of the matrix $A$.
  $\square$

 \section{Statement of the problem} 
 \subsection{General statement of the problem} 
  Let $R$ be a commutative domain, $F$ its field of quotients.
  Let a matrix $ {\mathcal M} \in R^{ N \times N} $ ($N=2^\nu$) be given, and let $ \alpha $ be the determinant of the largest nondegenerate (or empty) submatrix located in the upper left corner of the matrix $ {\mathcal M} $. For the case of an empty submatrix, we set $\alpha = 1 $.
  
  Let $ n = 2^p $, $ A \in R^{n \times n} $ be a matrix of rank 
  $ r $, $ r \le n $, and elements of the matrix $ A $ be surrounding  minors with respect to the minor $ \alpha $. In the case of an empty submatrix, we can take $ A = \mathcal M $.  Let $ det_{r}, det_{r-1},..,det_{1} $ be some sequence consisting of $r$ nested nondegenerate minors of matrix $A$.
  
  We want to obtain matrices $ L, U, M, W \in R^{n \times n} $ of rank $ n $, (L lower triangular, U - upper triangular), a   matrix $S \in {\bf S}_{wp}(F)$, of rank $ r $,
   with $r$ non-zero elements equal $(\alpha \det_{1})^{-1}$, $(det_{1} det_{2})^{-1}$, .., $(det_{r-1} det_r)^{-1}$,
  such that
 $$
 \left\{ \begin{array}{rcl} \alpha L S U = A  \\  L \widehat S M =   {\mathbf I}\\  W \widehat S U = {\mathbf I} \end{array}\right. .
 \eqno(1)
 $$
Here $det_{r}, det_{r-1},..,det_{1}$ is a sequence consisting of $r$ nested 
nondegenerate minors of the matrix $A$, which have sizes $r,r-1,..,1$, respectively.  ${\mathbf I}$ is the unit matrix and we denote  by $\widehat S$ a matrix 
$$\widehat S = \frac{1} {det_{r}}(\alpha S)^{Ext}=\frac{1} {det_{r}} (\alpha  S  +  \bar S).  $$   
  
As can be seen from (1), we want to find not only the triangular factors $L$ and $U$, but also their inverse matrices. These are matrices $\widehat S M $ and $W \widehat S$.
  
  Let us denote  
  $$  E =S^{\to 1}, I=EE^T, J=E^TE, \bar I={\bf I }-I, \bar J={\bf I }-J,\eqno(2)$$
  $E  \in {\bf S}_{p},\ I,J\in \bf  S_{d}$.  
  
  It should be noted that the matrices $ I $ and $ S $ have the same nonzero rows, and the matrices $ J $ and $ S $ have the same nonzero columns: 
$$ I S J = S,\  \ \bar I S  =0,\ \ S \bar J = 0.  $$
We define the properties of the matrices $ L $ and $ U $ as follows:
 $$L\bar I=\bar I ,\ \ \    \bar J U=\bar J.  \eqno(3)$$ 
\subsection{ Inverse and pseudo inverse matrices}
If matrix $A$ is invertible, then matrix 
$$
P= (1/ det_r^2) W S M
$$ 
 is the inverse matrix for $A$: $L^{-1}=\hat S M$,  $U^{-1}=W \hat S$,  $ S =det_r \hat S$, so  
 $$AP=LSU (1/ det_r^2) WSM = I.
 $$
 If matrix $A$ is not invertible, then one can easily check that two identities 
 $$ PAP=P \ \hbox{and } \ APA=A $$
hold. But, in the general case, identities $PA=(PA)^{T}$, $AP=(AP)^{T}$ do not hold. Thus, in this case, matrix $P$ is a pseudo-inverse, but is not the  Moore-Penrose generalized inverse for the matrix $A$.

\section{Construction of a dichotomous recursive factorization} 
 We want to describe a procedure that allows you to compute the LSU-factorization 
 $$
 (L, S, U, M, W, \alpha_r ) = {\bf LSU}(A, \alpha)
 $$
 in a recursive and dichotomous way.

 (1)   If $A=0$, then we assume that  $S=I=J=0$, $M=W=\alpha \bf I$,  ${\widehat S}=\alpha^{-1} \bf I$,
 $L=U=\bar I=\bar J=\bar S=\bf I$, $\alpha_r=\alpha$.

\noindent 
 (2) If $n=1$ ($A=[a], a \ne 0$), then  we assume that  $S=[(\alpha a)^{-1}]$, ${\widehat S}=[ a^{-2}]$, $L=U=M=W=[a]$, $ I= J= [1]$, $\bar I=\bar J=\bar S=[0]$, $\alpha_r=  a$.
 
\noindent
 (3) For $A\ne 0$ and $n>1$ we divide matrix $A$ into four equal blocks 
 $$
 A= \left[\begin{array}{cc}A_{11} & A_{12} \\ A_{21} & A_{22} \end{array}\right].  \eqno(4)
 $$
We can do the LSU-decomposition for block $A_{11}$:
 $$
\left\{ \begin{array}{rcl} \alpha L_{11}S_{11}U_{11}= \ A_{11} \\  L_{11}{\widehat S}_{11} M_{11} ={\mathbf I} \\  W_{11}{\widehat S}_{11} U_{11}={\mathbf I} \end{array}\right.. \eqno(5)
 $$ 
 Let $k=\, \mathbf{rank}(A_{11})$ and $det_{1}, det_{2}, .., det_k $ are the non-zero nested leading minors of $A_{11}$, that were found recursively, $\alpha_k =det_k$,   
 ${\widehat S}_{11}= \alpha_k ^{-1}(\alpha   S_{11}+ \bar S_{11})$
 and the non-zero elements of $S_{11}$ equal $(\alpha det_{1})^{-1}$, $(det_{1} det_{2})^{-1}$,...,$(det_{k-1} \alpha_k )^{-1}$.

 Then we can write the equality 
$    \left[\begin{array}{cc}A_{11} & A_{12} \\ A_{21} & A_{22} \end{array}\right]
  = $
  $$\left[\begin{array}{cc}L_{11} & 0 \\ 0 & {\mathbf I} \end{array}\right]   \left[\begin{array}{cc}  \alpha S_{11} & \frac { 1} {\alpha_k } A'_{12} \\   \frac 1  {\alpha_k } A'_{21} &    A_{22} \end{array}\right]   \left[\begin{array}{cc}U_{11} & 0 \\ 0 & {\mathbf I} \end{array}\right] . \eqno(6)$$ 
 We denote the new blocks
 $$
 A'_{12}=   { \alpha_k  }   {\widehat S}_{11} M_{11} A_{12},\ \  A'_{21} =  { \alpha_k   } A_{21} W_{11} {\widehat S}_{11}, \eqno(7)
 $$
 decompose the middle matrix  
 %
 $  \left[\begin{array}{cc}  \alpha S_{11} & {\frac{1}{\alpha_k} } A'_{12} \\  {\frac{1}{\alpha_k}} A'_{21} & A_{22} \end{array}\right]= 
 $
 $$
 \left[\begin{array}{cc}{\mathbf I} & 0 \\  {\frac{1}{\alpha_k}} A'_{21}S^{+}_{11} & {\mathbf I} \end{array}\right] 
 \left[\begin{array}{cc} 
 \alpha S_{11} & {\frac{\alpha}{\alpha_k} } A''_{12} \\ { \frac{1}{\alpha_k} } A''_{21} & {\frac{1}{\alpha_k}} A'_{22} \end{array}\right] 
\left[\begin{array}{cc}{\mathbf I} & { \frac{1}{\alpha_k}} S^{+}_{11} A'_{12}  \\ 0 & {\mathbf I} \end{array}\right] . \eqno(8) $$
 We denote 
 $$E_{11}=S_{11}^{\to 1},\ \ I_{11}= E_{11} E^T_{11},\ \ J_{11}= E^T_{11} E _{11}. $$
  
 As far as $\bar I_{11}  \widehat S_{11}=\bar S_{11}=\widehat S_{11}\bar J_{11}$, $S^+_{11}\bar I_{11}=0$ and $ S^+_{11}\widehat S_{11} =   \frac{\alpha} {\alpha_r} J_{11}$ we get    
 $$A''_{12}={  \frac{1}{\alpha \alpha_k}}\bar I_{11}A'_{12}=   {  \frac{1}{\alpha}}  \bar S_{11} M_{11} A_{12},
$$ $$ 
   A''_{21}={\frac{1}{\alpha \alpha_k}}A'_{21} \bar J_{11}=  { \frac{1}{\alpha}} A_{21} W_{11} \bar S_{11}. \eqno(9) $$

 For the lower right block we get
 $$ \alpha  \alpha_k ^{2} A_{22}= (A'_{21}J_{11}+A''_{21})S^{+}_{11} A'_{12}+  A'_{21} S^{+}_{11}  A''_{12}+ \alpha \alpha_k  A'_{22}$$
$$  A'_{22}=  { \frac{1}{\alpha \alpha_k} } ({\alpha\alpha_k ^{2}} A_{22} -  A'_{21}S^{+}_{11}  A'_{12}). \eqno(10)$$
Matrices $A''_{12}$ and $A''_{21}$   are matrices of surrounding minors with respect to minor $\alpha_k $. 
See the prove in \cite{2020ar} Theorem 8.2.  

Let  
 $$\left\{ \begin{array}{rcl} \alpha_k  L_{21}S_{21}U_{21}= A''_{21} \\  L_{21}\widehat S_{21}M_{21}= {\mathbf I} \\  W_{21}\widehat S_{21} U_{21}=   {\mathbf I} \end{array}\right.   \hbox{ and }  
 \left\{ \begin{array}{rcl} \alpha_k  L_{12}S_{12}U_{12}= A''_{12} \\  L_{12}\widehat S_{12}M_{12}= {\mathbf I} \\  W_{12}\widehat S_{12} U_{12}= {\mathbf I}    \end{array}\right.$$ 
 be LSU-decomposition of blocks $A''_{21}$ and  $A''_{12}$. 

 Let $A'$ be submatrix of $A$ which is fixed with non zero rows of  the 
 matrix $ diag(I_{11}, I_{12}, I_{21})$ and non zero columns of the 
 matrix  $ diag(J_{11}, J_{21}, J_{12})$.
 Submatrices $A''_{12}$, $A''_{21}$ and the submatrix, which corresponds to 
 minor $\alpha_k $, do not have common nonzero rows and columns, so the 
 sequence of nested non zero minors of submatrix $A'$ can be selected in 
 different ways.

Let $rank(A''_{21})=l_1$ and $rank(A''_{12})=m_1$, then the rank of submatrix  $A'$ is equal $s=k+m_1+l_1$. Suppose that we have obtained the following sequences of nested minors: $det_{1},...,det_{k},  ,...,det_{l}$, $(l=l_1+k)$,  for the block 
$$ \left[ \begin{array}{c} A_{11} \\ A_{21}  \end{array}\right]$$
and $det_{1},...,det_{k},det'_{k+1}...,det'_m$, $m= m_1+k $,  for the block 
$$ \left[ \begin{array}{cc} A_{11} & A_{12}  \end{array}\right].$$

 For the matrix $A'$ we can set the following sequences of nested minors:
$$det_{1},...,det_{k},...,det_{l},det_{l+1},,...,det_{s}, $$
with 
$$det_{l+i}=\lambda det'_{k+i },\ \ i=1,2,..,m_1,\ (\lambda= det_{l}/det_{k}),  $$
  in particular, $det_{s}=\lambda det'_{m} $. 
  
  We denote $\alpha_l  =det_{l}, \ \alpha_s=det_{s}$, $J^{\lambda}_{12}= \lambda  J_{12}+\bar J_{12} $, $I^{\lambda}_{12}= \lambda  I_{12}+\bar I_{12}$,  
  $$   \widehat S_{21}  =\frac {\alpha_k   S_{21}  +   \bar S_{21}} {\alpha_m} , \ 
  \widehat S_{12}  =\frac {\alpha_k   S_{12}  +   \bar S_{12}} {\alpha_l } , 
  $$ $$\ 
  L^{\sim}_{12}=  L_{12}  I^{\lambda}_{12},\ \ U^{\sim}_{12}= J^{\lambda}_{12} U_{12}, 
  \eqno(11)$$
   $$ S^{\sim}_{12}=\lambda^{-2} S _{12}, \ \  {\widehat S}^{\sim}_{12}= I^{\lambda^{-2}}_{12} {\widehat S} _{12}, 
$$ $$   
   \ \ 
  M^{\sim}_{12}=J^{\lambda}_{12} M_{12},  \ \ W^{\sim}_{12}=  W_{12}I^{\lambda}_{12}, \eqno(12)$$
  Then the last system can be written as follows:
 $$\left\{ \begin{array}{rcl} (\lambda \alpha_k ) L^{\sim}_{12} S^{\sim}_{12}U^{\sim}_{12}= \lambda A''_{12} \\  L^{\sim}_{12}{\widehat S}^{\sim}_{12}M^{\sim}_{12}= {\mathbf I}  \\  W^{\sim}_{12}{\widehat S}^{\sim}_{12} U^{\sim}_{12}=   {\mathbf I}    \end{array}\right.,$$ 
and the   matrix equation 
  $ \left[\begin{array}{cc} \alpha  S_{11} & { \frac{\alpha}{\alpha_k} } A''_{12} \\ {  \frac{1}{\alpha_k} } A''_{21} & { \frac{1}{\alpha_k} } A'_{22} \end{array}\right]=
$  $$  
  \left[\begin{array}{cc}L^{\sim}_{12} & 0 \\ 0 & L_{21} \end{array}\right]   \left[\begin{array}{cc} 
  \alpha S_{11} & \alpha S^{\sim}_{12} \\ \alpha S_{21} & {  \frac{1}{\alpha_k} } A''_{22} \end{array}\right] 
  \left[\begin{array}{cc}U_{21} & 0 \\ 0 & U^{\sim}_{12} \end{array}\right], \eqno(13)$$
 is true, where we denote
  $A''_{22}=  
  $ $$ {\widehat S}_{21}M_{21}A'_{22}  W^{\sim}_{12} {\widehat S}^{\sim}_{12}={\widehat S}_{21}M_{21}A'_{22}  W_{12}  I^{\lambda^{-1}}_{12} {\widehat S}_{12}
 \eqno(14)$$
 and use the following equation:  $I_{12}  I_{11}=0$,    $J_{11}  J_{21}=0$ and $L_{12} S_{11} U_{21}=S_{11}$. To check the last equation we can write each matrices as follows: $L_{12}=  L_{12} I_{12}+ \bar I_{12}$, $S_{11}=  I_{11}  S_{11} J_{11}$,   $U_{21}=   J_{21}  U_{21}+  \bar J_{21}$. 

 The middle matrix 
  $ \left[\begin{array}{cc} \alpha S_{11} & \alpha  S^{\sim}_{12} \\ \alpha S_{21} & { \frac{1}{\alpha_k} } A''_{22} \end{array}\right] 
  $
  can be decomposed in two ways:
  $$   \left[\begin{array}{cc}{\mathbf I}  & 0 \\   
  \frac {A''_{22}S^{\sim +}_{12}} {\alpha \alpha_k} & {\mathbf I}  \end{array}\right]    \left[\begin{array}{cc} \alpha S_{11} & \alpha S^{\sim}_{12} \\ \alpha S_{21} & {\frac{\alpha}{\alpha_s}}  A'''_{22} \end{array}\right]    \left[\begin{array}{cc}{\mathbf I}  &  \frac {  S^{+}_{21} A''_{22} \bar J_{12}} {\alpha \alpha_k}  \\ 0 & {\mathbf I}  \end{array}\right] 
  $$
 or
  $$
  \left[\begin{array}{cc}{\mathbf I}  & 0 \\   \frac{\bar I_{21}  A''_{22}S^{\sim +}_{12}}  { \alpha \alpha_k } & {\mathbf I}  
  \end{array}\right] \left[\begin{array}{cc} 
  \alpha S_{11} &  \alpha  S^{\sim}_{12} \\  \alpha S_{21} & { \frac{\alpha}{\alpha_s}}  A'''_{22} 
  \end{array}\right]    \left[\begin{array}{cc}
  {\mathbf I}  & \frac {  S^{+}_{21} A''_{22}}  { \alpha \alpha_k }   \\ 0 & {\mathbf I}  \end{array}\right]. 
  \eqno(15)$$
  
 We use the following equations $S^{+}_{12} S_{11}=0$, $S_{11} S^{+}_{21}=0$, 
and for the lower right block we get
 $$   {1  \over  \alpha_k } A''_{22}=   {1  \over  \alpha_k ^{2} }  A''_{22} J_{12} +   { \frac{1}{\alpha_k^{2}} }  I_{21}A''_{22}\bar J_{12}+ { \frac{\alpha}{\alpha_s}}    A'''_{22}, $$
or
 $$ {1  \over  \alpha_k   } A''_{22}=  { \frac{1}{\alpha_k} } I_{21}  A''_{22} +  { \frac{1}{\alpha_k}  } \bar  I_{21}A''_{22} J_{12}+ { \frac{\alpha}{\alpha_s}}   A'''_{22}.$$
 Both of these expressions give the same value for $ A'''_{22}$:
 
  $$   A'''_{22}=   {\alpha_s  \over  \alpha\alpha_k   } \bar I_{21}A''_{22}\bar J_{12}= {1 \over  \alpha_k ^{2} \alpha  } \bar S_{21}M_{21} A'_{22} W_{12} \bar S_{12}. \eqno(16)$$
 Further we will use the second decomposition.
 
 The  matrix $A'''_{22}$ is the matrix of surrounding minors of $A$ with respect to minor $\alpha_s$.
See the prove in \cite{2020ar} .

 Let  
 $$ \left\{ \begin{array}{rcl} \alpha_s L_{22} S_{22} U_{22}=  A'''_{22} \\  L_{22}\widehat S_{22} M_{22}=  {\mathbf I}  \\  W_{22} \widehat S_{22} U_{22} =   {\mathbf I}     \end{array}\right. $$
 be a decomposition of the matrix $ A'''_{22}$, $\widehat S_{22}= \frac {1}{\alpha_r} (\alpha_s  S_{22}  +   \bar S_{22}) $,  
 then we can write the equality 
 $$    \left[\begin{array}{cc}S_{11} &   S^{\sim}_{12} \\   S_{21} &   \frac  {A'''_{22}}{\alpha_s} \end{array}\right]   = \left[\begin{array}{cc}{\mathbf I}  & 0 \\ 0 & L_{22} \end{array}\right]   \left[\begin{array}{cc}   S_{11} &    S^{\sim}_{12} \\ S_{21} &  S_{22} \end{array}\right]    \left[\begin{array}{cc}{\mathbf I}  & 0 \\ 0 & U_{12} \end{array}\right]. \eqno(17)$$
  To prove this equality  we can write   matrices $L_{22}$, $U_{22}$, $S_{12}$, $S_{12}$ as follows 
 $ L_{22}=   L_{22} I_{22}+  \bar I_{22},\  U_{22}= J_{22} U_{22} + \bar   J_{22},$  
$ S_{21}=  I_{21}  S_{21}  J_{21},\ S_{12}=   I_{12}  S_{12}  J_{12},$ 
 and check the equations 
 $$L_{22} S_{21}=S_{21} \hbox{  and  }   S^{\sim}_{12}U_{22}=S^{\sim}_{12}.$$
As a result of the  sequence (6), (8), (13), (15),(17) of decompositions we obtain the LSU-decomposition of the matrix A in the form
  $$    \left[\begin{array}{cc}A_{11} & A_{12} \\ A_{21} & A_{22} \end{array}\right]  = 
  \alpha L  S  U, 
$$ $$  
  \ \ S=
  \left[\begin{array}{cc} S_{11} & S^{\sim}_{12} \\ S_{21} & S_{22} \end{array}\right]=
    \left[\begin{array}{cc} S_{11} & \lambda^{-2} S_{12} \\ S_{21} & S_{22} \end{array}\right], 
  \eqno(18)$$ 
 with $S^{\sim}_{12}=\lambda^{-2}  S_{12}$  and such matrices $L$ and $U$:
 $$
L=\left[\begin{array}{cc}L_{11} & 0 \\ 0 & {\mathbf I} \end{array}\right]  \left[\begin{array}{cc}{\mathbf I} & 0 \\ \frac 1{\alpha_k } A'_{21}S^{+}_{11} & {\mathbf I} \end{array}\right]    \left[\begin{array}{cc}L^{\sim}_{12} & 0 \\ 0 & L_{21} \end{array}\right]  
$$ $$
\times \left[\begin{array}{cc}{\mathbf I} & 0 \\ \frac {\lambda^2}{\alpha  \alpha_k }\bar {\mathbf I}_{21} A''_{22}S^{+}_{12} & {\mathbf I} \end{array}\right]\left[\begin{array}{cc}{\mathbf I} &  0 \\ 0 &  L_{22} \end{array}\right],
$$
 $$U= \left[\begin{array}{cc}{\mathbf I} &  0 \\ 0 &  U_{22} \end{array}\right]    \left[\begin{array}{cc}{\mathbf I} &  \frac 1{\alpha  \alpha_k }S^{+}_{21}A''_{22} \\ 0 & {\mathbf I} \end{array}\right]   \left[\begin{array}{cc}U_{21} & 0 \\ 0 & U^{\sim}_{12} \end{array}\right]  
$$ $$ 
  \times    \left[\begin{array}{cc}{\mathbf I} & \alpha_k ^{-1} S^{+}_{11} A'_{12} \\ 0 & {\mathbf I} \end{array}\right]   \left[\begin{array}{cc}U_{11} & 0 \\ 0 & I \end{array}\right]. $$
 After multiplying the matrices on the right side, we get 
 $$ L= \left[\begin{array}{cc}   L_{1}  & 0 \\ L_{3} & L_{4} \end{array}\right]=\left[\begin{array}{cc}   L_{11}L^{\sim}_{12} & 0 \\ L_{3} & L_{21}L_{22} \end{array}\right], 
$$ $$ 
 \ \   U =\left[\begin{array}{cc} U_{1} & U_{2} \\   0 & U_{4} \end{array}\right]=  \left[\begin{array}{cc}U_{21}U_{11} & U_{2} \\   0 & U_{22}U^{\sim}_{12} \end{array}\right], \eqno(19)$$
with
 $$ U_{2}= \frac 1{\alpha_k } U_{21} S^{+}_{11} A'_{12}+ \alpha^{-1}\alpha_k ^{-1}  S^{+}_{21}A''_{22} U^{\sim}_{12}  = $$
 $$=\alpha_k  ^{-1} J_{11} M_{11} A_{12}+  \alpha^{-1}\alpha_l ^{-1} J_{21} M_{21}A'_{22},  \eqno(20)$$
  $$L_{3}=\alpha_k ^{-1} A'_{21} S^{+}_{11}L^{\sim}_{12}+{\alpha^{-1}  \alpha_k^{-1}  } L_{21} \bar {\mathbf I}_{21} A''_{22}S^{\sim +}_{12} =$$
  $$=\alpha_k ^{-1} A_{21}W_{11} I_{11} + (\alpha \alpha_m \alpha_k) ^{-1} \bar S_{21} M_{21} A'_{22} W_{12} I_{12}, \eqno(21)$$
   $$ {\widehat S}=\alpha_r^{-1}(\alpha S +\bar S), \eqno(22)$$
 $$M= {\widehat S}^{-1}  \left[\begin{array}{cc}   L_{1}^{-1}  & 0 \\ -L_{4}^{-1}L_{3}L_{1}^{-1} & L_{4}^{-1} \end{array}\right], 
$$ $$ 
 \ 
 W =  \left[\begin{array}{cc}   U_{1}^{-1}  &  -U_{1}^{-1}U_{2} U_{4}^{-1} \\ 0 & U_{4}^{-1} \end{array}\right] 
  {\widehat S}^{-1}.
 $$

We can use expressions $L_{ij}^{-1}=\widehat S_{ij} M_{ij}$ and $U_{ij}^{-1}=W_{ij} \widehat S_{ij} $ to simplify the calculation of matrices $M$ and $W$.
Let us denote  
   $$U_{2}'=W_{11} \widehat S_{11} W_{21} \widehat S_{21} U_2, \ \
   L_{3}'= L_3 I^{\lambda^{-1}}_{12} {\widehat S}_{12} M_{12} {\widehat S}_{11} M_{11},$$
then we get
 $$M= {\widehat S}^{-1}
  \left[\begin{array}{cc} I^{\lambda^{-1}}_{12}{\widehat S}_{12} M_{12} {\widehat S}_{11} M_{11}  & 0 \\ 
0 &  {\widehat S}_{22} M_{22} {\widehat S}_{21} M_{21} 
 \end{array}\right]
 $$ $$ \times
  \left[\begin{array}{cc} {\mathbf I}  & 0 \\ -L_{3}' & {\mathbf I}  \end{array}\right],    \eqno(23)
$$
  $$W=  \left[\begin{array}{cc}  {\mathbf I} & -U_{2}'  \\ 0 & {\mathbf I} \end{array}\right] 
  $$ $$ \times
  \left[\begin{array}{cc}  W_{11} \widehat S_{11} W_{21} \widehat S_{21}  & 0   \\
  0 &  W_{12} \widehat S_{12}J^{\lambda^{-1}}_{12} W_{22} \widehat S_{22} \end{array}\right] {\widehat S}^{-1}, \eqno(24)$$
      $$ 
     \widehat S^{-1}=  \frac {\alpha_r} {\alpha} (S^{+}+ \alpha \bar S^{T})
$$ $$     
     = 
    \frac {\alpha_r} {\alpha} \left[\begin{array}{cc}  S_{11}^{+}+\alpha \tilde S_{11}^{T} &    S_{21}^{+}+\alpha \tilde   S_{21}^{T} \\  \lambda^{2} S_{12}^{+}+\alpha \tilde S_{12}^{T} &   S_{22}^{+}+ \alpha \tilde S_{22}^{T} \end{array}\right]. \eqno (25)
      $$ 
      Here we denote the blocks of the matrix $\bar S$  by $\tilde S_{ij}$, $(i,j=1,2)$.
      
 As far as $ S_{11}^{+}  S_{12}=0 $  and $\tilde S_{11}^{T}  S_{12}=0$, we get 
 $$(  S_{11}^{+}+\alpha \tilde S_{11}^{T}) I^{ \lambda^{-1}}_{12}{\widehat S}_{12} 
 = (  S_{11}^{+}+\alpha \tilde S_{11}^{T}){\bar S}_{12}\frac 1 {\alpha_m}
$$
and 
$$  (  S_{11}^{+}+\alpha \tilde S_{11}^{T}){\widehat S}_{12} M_{12} \widehat S_{11} M_{11} 
$$ $$
=
 (E_{11}^{T}+ \tilde S_{11}^{T}){\bar S}_{12} M_{12} ( E_{11}+   \bar S_{11})M_{11} \frac {\alpha} {\alpha_m \alpha_k }. $$
 
 From such identities we get:
 $$    (  S_{12}^{+}+\alpha_k  \tilde S_{12}^{T}){\widehat S}_{12} M_{12} \widehat S_{11} M_{11} 
$$ $$ 
 =
 (E_{12}^{T}+  \tilde S_{12}^{T})(\alpha_l  E_{12}+ \alpha {\bar S}_{12}) M_{12}  \bar S_{11} M_{11} \frac 1 {\alpha_m \alpha_k }, $$ 
$$   (  S_{21}^{+}+\alpha_k  \tilde S_{21}^{T}){\widehat S}_{22} M_{22} \widehat S_{21} M_{21} 
$$ $$
=
 (E_{21}^{T}+  \tilde S_{21}^{T}){\bar S}_{22} M_{12}(\alpha_k  E_{21}+ \alpha \bar S_{21}) M_{21} \frac 1 {\alpha_l  \alpha_r}, $$
  $$   (  S_{22}^{+}+\alpha_s \tilde S_{22}^{T}){\widehat S}_{22} M_{22} \widehat S_{21} M_{21} 
$$ $$  
  =
 (E_{22}^{T}+  \tilde S_{22}^{T})(\alpha_s   E_{22}+ \alpha {\bar S}_{22}) M_{22}  \bar S_{21} M_{21} \frac 1 {\alpha_l  \alpha_r}. 
 $$ 
Therefore, if we denote  $(E_S+\bar S)^T$
 $$
\left[\begin{array}{cc}  E'_{11}  &  E'_{12} \\ E'_{21} & E'_{22}  \end{array}\right]= 
\left[\begin{array}{cc}   E_{11}^{T}+ \tilde S_{11}^{T}  & E_{21}^{T}+ \tilde S_{21}^{T} \\
   E_{12}^{T}+\tilde S_{12}^{T} & E_{22}^{T}+\tilde S_{22}^{T} \end{array}\right], 
$$
$$
 \left[\begin{array}{cc}  S'_{11}  &  S'_{12} \\ S'_{21} & S'_{22}  \end{array}\right]=
 \left[\begin{array}{cc} \alpha E_{11}+\alpha \bar S_{11}  & \alpha_k  E_{21}+\alpha \bar S_{21} \\
 \alpha_l  E_{12}+\alpha \bar S_{12} &\alpha_s E_{22}+\alpha \bar S_{22} \end{array}\right], 
 $$
 $$ L_{3}'= L_3 I^{\lambda^{-1}}_{12} {\widehat S}_{12} M_{12} {\widehat S}_{11} M_{11}, \ \ 
      U_{2}'=W_{11} \widehat S_{11} W_{21} \widehat S_{21} U_2,$$ 
then,   based on  (23) and  (25), we get 
$$ M= 
\left[\begin{array}{cc} E'_{11} \bar S_{12} M_{12} S'_{11} M_{11}   &
E'_{12} \bar S_{22} M_{22} S'_{21} M_{21}   \\
E'_{21} S'_{12} M_{12} \bar S_{11} M_{11}   &
E'_{22}  S'_{22} M_{22} \bar S_{21} M_{21}   
 \end{array}\right]
$$ $$\times  
 \left[\begin{array}{cc}   \frac{1}{\alpha_m \alpha_k }   &  0  \\ -\frac{1}{\alpha_l  \alpha_r}  L_{3}' &  \frac{1}{\alpha_l  \alpha_r}   \end{array}\right] \frac{\alpha_r}{\alpha}. \eqno (26)$$
 Similarly, it is easy to show that starting from (24) and  (25), we get 
 $$W= \frac{\alpha_r}{\alpha}
 \left[\begin{array}{cc}  \frac{1}{\alpha_k  \alpha_l }  &  - \frac{1}{\alpha_m \alpha_r} U'_2  \\ 0 &  \frac{1}{\alpha_m \alpha_r}  \end{array}\right]
$$ $$ \times 
  \left[\begin{array}{cc}  W_{11} S'_{11} W_{21} \bar S_{21} E'_{11}    &
    W_{11} \bar S_{11} W_{21}  S'_{21} E'_{12}  \\
    W_{12} S'_{12} W_{22} \bar S_{22} E'_{21}   & 
    W_{12} \bar S_{12} W_{22} S'_{22} E'_{22}  
 \end{array}\right]. \eqno (27)$$

 In the   expressions (18) - (27),  the LSU-decomposition of the matrix $ A $ are given and the matrices $ W $ and $ M $  that satisfy the conditions $ L\hat S M = {\mathbf I}  $ and $ W\hat S U = {\mathbf I}  $ are obtained.
 
 We proved the correctness of the following recursive algorithm.
\section{ Algorithm of LSU factorization  }

{\bf Input:}  matrix $A\in R^{n\times n}, n=2^p, p>0$, $\alpha=1$. 

\noindent
{\bf Output:} matrices $L, S, U, M, {\widehat S}, W$, and number $\alpha_r$.
      $$(L, S, U, M, {\widehat S}, W, \alpha_r)=\, \mathbf{LSU}(A, \alpha).$$

      \noindent
 {\bf (1)}   If ($A=0$)  then 
 
  $\{$  $\alpha_r = \alpha;  $ 
 $S =I=J= 0; $
  $ L =U = \bar I = \bar J= \bar S  = I; $ 
    
               $ M = W =\widehat  S = \alpha I;$ $\}$
              
\noindent   
 {\bf (2)} If ($n=1$ $\&$ $A=[a]$  $\&$ $a\ne 0$)  then 
 
  $\{ \alpha_r =a;\ L = U = M = W = [a];
                S = [(\alpha  a)^{-1}]; $
                
               $ \widehat  S=[a^{-2}];$             
                $J=I=[1];
                \bar I= \bar J = \bar S = [0];  \}$  
  
 \noindent
 {\bf (3)} If ($n\ge 2$ $\&$ $A \ne 0$) then   
  $\bigg\{$
 $A= \left[\begin{array}{cc}A_{11} & A_{12} \\ A_{21} & A_{22} \end{array}\right] $; 
 
$$ (3.1) \hspace{6mm}   (L_{11}, S_{11}, U_{11}, M_{11}, W_{11}, \alpha_k  )=\, \mathbf{LSU}(A_{11}, \alpha);\hspace{16mm}$$ 

      $A_{12}^0= M_{11}  A_{12};$ 

     $ A_{12}^1= \alpha_k    \widehat  S_{11}    A_{12}^0;$
      
      $A_{12}^2= \bar S_{11}    A_{12}^0 / \alpha;$
      
      $A_{21}^0=A_{21}  W_{11};$  
      
      $A_{21}^1 =\alpha_k     A_{21}^0   \widehat  S_{11};$
      
      $A_{21}^2= A_{21}^0    \bar S_{11} / \alpha;$

$$(3.2) \hspace{6mm}   (L_{21}, S_{21}, U_{21}, M_{21}, W_{21}, \alpha_l  )=\, \mathbf{LSU}(A_{21}^2, \alpha_k );\hspace{16mm}$$
$$(3.3)  \hspace{6mm}   (L_{12}, S_{12}, U_{12}, M_{12}, W_{12}, \alpha_m )=\, \mathbf{LSU}(A_{12}^2, \alpha_k );\hspace{16mm}$$     
 
        $\lambda=  {\alpha_l  \over \alpha_k }; \alpha_s=\lambda  \alpha_m;$  
 
         $A_{22}^0= A_{21}^1  S_{11}^{+}  A_{12}^1;$
         
          $A_{22}^1= ( \alpha \alpha_k ^2  A_{22} - A_{22}^0 )/(\alpha \alpha_k );$
         $ A_{22}^2=\bar S_{21}  M_{21}  A_{22}^1  W_{12}   \bar  S_{12};$
         
          $A_{22}^3=A_{22}^2/(\alpha_k ^2 \alpha);$      
$$(3.4) \hspace{6mm}   (L_{22}, S_{22}, U_{22}, M_{22}, W_{22}, \alpha_r )=\, \mathbf{LSU}(A_{22}^3, \alpha_s);\hspace{16mm}$$ 
 
                $J_{12}^{\lambda}=  \lambda   J_{12}  + \bar J_{12};\ \ 
                I_{12}^{\lambda}= \lambda I_{12}   + \bar I_{12};$  
                
               $  L_{12}^{\sim} = L_{12}  I_{12}^{\lambda};\ \ 
                 U_{12}^{\sim} = J_{12}^{\lambda}  U_{12};$        
   
        $U_{2}=J_{11}  M_{11}  A_{12}/\alpha_k  + J_{21}  M_{21}   A_{22}^1 /(\alpha_l   \alpha);$
        
        $ L_{3}= A_{21}  W_{11}  I_{11}/\alpha_k  +\bar S_{21}  M_{21}   A_{22}^1  W_{12}  I_{12}/(\alpha_m   \alpha_k    \alpha);$
$$
L= \left[\begin{array}{cc}   L_{11}  L^{\sim} _{12} & 0 \\ L_{3} & L_{21}L_{22} \end{array} \right];  \ \ 
S=\left[\begin{array}{cc}   S_{11} &   \lambda^{-2}  S_{12} \\    S_{21} &   S_{22} \end{array}\right]; 
$$ $$
\ \ 
 U =  \left[\begin{array}{cc}U_{21}U_{11} & U_{2} \\   0 & U_{22}   U^{\sim}_{12} \end{array}\right]; 
$$
 $
\widehat S= \alpha (\alpha_r)^{-1}S+ (\alpha_r)^{-1}\bar S; \ \ E=S^{\to 1};
$ 
%
\\
$ L_{3}'= L_3 I^{\lambda^{-1}}_{12} {\widehat S}_{12} M_{12} {\widehat S}_{11} M_{11};$ \\
$U_{2}'=W_{11} \widehat S_{11} W_{21} \widehat S_{21} U_2;$ \\
      
 $ (E+\bar S)^T= $ \\
 $$ =
\left[\begin{array}{cc}  E'_{11}  &  E'_{12} \\ E'_{21} & E'_{22}  \end{array}\right]= 
\left[\begin{array}{cc}   E_{11}^{T}+ \tilde S_{11}^{T}  & E_{21}^{T}+ \tilde S_{21}^{T} \\
E_{12}^{T}+\tilde S_{12}^{T} & E_{22}^{T}+\tilde S_{22}^{T} \end{array}\right]; 
$$
$$
 \left[\begin{array}{cc}  S'_{11}  &  S'_{12} \\ S'_{21} & S'_{22}  \end{array}\right]=
 \left[\begin{array}{cc} \alpha E_{11}+\alpha \bar S_{11}  & \alpha_k  E_{21}+\alpha \bar S_{21} \\
 \alpha_l  E_{12}+\alpha \bar S_{12} &\alpha_s E_{22}+\alpha \bar S_{22} \end{array}\right];
 $$
  $$ M=\frac{1}{\alpha} 
\left[\begin{array}{cc} E'_{11} \bar S_{12} M_{12} S'_{11} M_{11}   &
E'_{12} \bar S_{22} M_{22} S'_{21} M_{21}   \\
E'_{21} S'_{12} M_{12} \bar S_{11} M_{11}   &
E'_{22}  S'_{22} M_{22} \bar S_{21} M_{21}   
 \end{array}\right]
 $$ $$ \times
 \left[\begin{array}{cc}   \frac{\alpha_r}{\alpha_m \alpha_k } I  &  0  \\ -\frac{1}{\alpha_l}  L_{3}' 
 &  \frac{1}{\alpha_l} I  \end{array}\right]; $$
 $$W= \frac{1}{\alpha}
 \left[\begin{array}{cc}  \frac{\alpha_r}{\alpha_k  \alpha_l } I &  - \frac{1}{\alpha_m} U'_2  \\ 0 &  \frac{1}{\alpha_m} I \end{array}\right]
 $$ $$ \times
  \left[\begin{array}{cc}  W_{11} S'_{11} W_{21} \bar S_{21} E'_{11}    &
    W_{11} \bar S_{11} W_{21}  S'_{21} E'_{12}  \\
    W_{12} S'_{12} W_{22} \bar S_{22} E'_{21}   & 
    W_{12} \bar S_{12} W_{22} S'_{22} E'_{22}  
 \end{array}\right].
\bigg\}
      $$

 %
 
%


\section{Complexity}
 
 Let $\gamma$ and $\beta$ be constants, $3\geq\beta>2$, and let
  ${\mathcal M}(n)= \gamma n^{\beta} + o(n^{\beta})$ be the number of  arithmetic
  operations in one $n\times n$ matrix multiplication.
 
  \subsection{The number of arithmetic operations in the numeric field}

\begin{theorem}
 The LSU factorithation algorithm has the complexity of matrix multiplication.
\end{theorem}
{\bf Proof.} 

The total number of matrix multiplications for calculating the factors $L,S,U$ is 13, and for calculating the matrices $M$ and $W$ is 16. The total number of recursive calls is 4. We do not consider permutation matrix multiplications such as $S, E, J, I$, since their complexity does not exceed $n^2$.

 Let $T_2$ be the number of operations for LSU factorization of the second order matrix.
  Therefore  we get the following recurrent equality for complexity
$$
t(n)=4 t(n/2)+ 29 {\mathcal M}(n/2), t(2)=T_2.
$$

After summation from  $n=2^k$ to $2^1$ we obtain
$$
29 \gamma(4^0 2^{\beta(k-1)} + \ldots + 4^{k-2}2^{\beta *1})+4^{k-1} T_2
$$ $$ 
  = 29 \gamma\frac{n^{\beta}-2^{\beta-2}n^2}{2^{\beta}-4} +
\frac{T_2}{4}n^2.
$$
Therefore the complexity of the LSU algorithm  is
 $\sim  \sigma n^{\beta}$,  
 $\sigma=\frac{29\gamma}{2^{\beta}-4}$.

 $\square$
 
If we take into account the rank of the matrix, then, similar to other
 algorithms, this estimate is reduced to 
 $\sim  \sigma r n^{\beta-1}$.

 \section{Examples}
  \subsection{Detailed description of an example of matrix factorization}
 
We give below an example of a LSU factorization of a $4\times 4$ matrix $A$:

\noindent 
$$A=\left[\begin{array}{cccc}0 &  0 &  3 &  0 \\ 2 &  0 &  1 &  0 \\ 0 &  0 &  0 &  0 \\ 1 &  4 &  0 &  1 \end{array}\right].$$

 First step: factorization of the upper left block.\\
$\mathbf{LSU}\Big ( 
A_{11}=\left[\begin{array}{cc}0 &  0 \\ 2 &  0 \end{array}\right], 1 \Big ):\\ 
L_{11}=\left[\begin{array}{cc}1 &  0 \\ 0 &  2 \end{array}\right];     
S_{11}=\left[\begin{array}{cc}0 &   0 \\ 1/2 &  0 \end{array}\right]; 
U_{11}=\left[\begin{array}{cc}2 &  0 \\ 0 &  1 \end{array}\right]; $ \\
$M_{11}=\left[\begin{array}{cc}0 & \ 2 \\ 2 & \ 0 \end{array}\right];
 W_{11}=\left[\begin{array}{cc}0 &  2 \\ 2 &  0 \end{array}\right];$ 
  $ \widehat S_{11}=\left[\begin{array}{cc}0 &  \frac 1 2 \\ \frac 1 4 &  0 \end{array}\right];  
\\ 
  \bar S_{11}=\left[\begin{array}{cc}0 &  1 \\ 0 &  0  \end{array}\right];$ 
$ I_{11}=\left[\begin{array}{cc}0 &  0 \\ 0 &  1  \end{array}\right];
J_{11}=\left[\begin{array}{cc}1 &  0 \\ 0 &  0  \end{array}\right]; \\
  \alpha_k =2; \ \alpha =1; $ \\
 $A_{12}^0
 =\left[\begin{array}{cc}2 &  0 \\ 6 &  0 \end{array}\right];
  A_{12}^1=  \left[\begin{array}{cc}6 &  0 \\ 1 &  0 \end{array}\right];$
   \\   
      $A_{21}^0
      = \left[\begin{array}{cc}0 &  0 \\ 8 &  2 \end{array}\right];
      A_{21}^1 =  \left[\begin{array}{cc}0 &  0 \\ 1 &  8 \end{array}\right];
    $
    
   Second step: factorization of the lower left block. \\ 
 $\mathbf{LSU}\Big (A_{21}^2  = \left[\begin{array}{cc}0 &  0 \\ 0 &  8 \end{array}\right], \alpha_k \Big );$\\
$ L_{21}= \left[\begin{array}{cc}1 &  0 \\ 0 &  8 \end{array}\right];$ 
$ S_{21}= \left[\begin{array}{cc}0 &  0 \\ 0 &  \frac 1{16} \end{array}\right]; U_{21}= \left[\begin{array}{cc}1 &  0 \\ 0 &  8 \end{array}\right];  \\
M_{21}= \left[\begin{array}{cc}8 &  0 \\ 0 &  8 \end{array}\right]; W_{21}= \left[\begin{array}{cc}8 &  0 \\ 0 &  8 \end{array}\right]; \ $ $ \widehat S_{21}= \left[\begin{array}{cc}\frac 1 8  &  0 \\ 0 & \frac 1 {64} \end{array}\right];   $
$ \bar S_{21}= \left[\begin{array}{cc}1 &  0 \\   0 &  0 \end{array}\right]. $

Third step: factorization of the upper right block.\\ 
$\mathbf{LSU}\Big ( A_{12}^2  =  \left[\begin{array}{cc}6 &  0 \\ 0 &  0 \end{array}\right], \alpha_k\Big ); \\
L_{12}= \left[\begin{array}{cc}6 &  0 \\ 0 &  1 \end{array}\right]; S_{12}= \left[\begin{array}{cc} \frac 1 {12} &  0 \\ 0 &  0 \end{array}\right] ;  U_{12}= \left[\begin{array}{cc}6 &  0 \\ 0 &  1 \end{array}\right];  
\\
M_{12}= \left[\begin{array}{cc}6 &  0 \\ 0 &  6 \end{array}\right]; W_{12}= \left[\begin{array}{cc}6 &  0 \\ 0 &  6 \end{array}\right]; 
\widehat S_{12}= \left[\begin{array}{cc}\frac 1 {36} &  0  \\ 0 &   \frac 1 6 \end{array}\right]; \\
\bar S_{12}= \left[\begin{array}{cc} 0 &  0  \\ 0 & 1 \end{array}\right];$ 
$I_{12}=J_{12}= \left[\begin{array}{cc} 1 &  0  \\ 0 & 0 \end{array}\right];
I_{21}=J_{21}= \left[\begin{array}{cc} 0 &  0  \\ 0 & 1 \end{array}\right];$
$L_{12}^{\sim}=U_{12}^{\sim}= \left[\begin{array}{cc}24 &  0 \\ 0 &  1 \end{array}\right];
 \alpha_l=8; 
   \alpha_m=6; \alpha_s=24; \lambda=4. $ 
$A_{22}^0= A_{21}^1  S_{11}^{+}  A_{12}^1= \left[\begin{array}{cc}0 & 0 \\ 2 &  0 \end{array}\right]; 
A_{22}^1= \frac{ \alpha \alpha_k ^2  A_{22} - A_{22}^0 }{\alpha \alpha_k } = \left[\begin{array}{cc}0 &  0 \\ -1 & 2 \end{array}\right];$
$ A_{22}^2=  \left[\begin{array}{cc}0 &  0 \\ 0 &  0 \end{array}\right].$
   
 Fourth step: factorization of the lower right block.\\
$\mathbf{LSU}\Big (A_{22}^3= \left[\begin{array}{cc}0 &  0 \\ 0 &  0 \end{array}\right],\alpha_s\Big );$
  \\     %
       $L_{22}=U_{22}=\left[\begin{array}{cc}1 &  0 \\ 0 &  1 \end{array}\right]; 
       S_{22}=\left[\begin{array}{cc}0 &  0 \\ 0 &  0 \end{array}\right];        
 $ 
 $M_{22}=W_{22}= \left[\begin{array}{cc}24 &  0  \\ 0 &   24 \end{array}\right]; \widehat S_{22}=\left[\begin{array}{cc}1/24 &  0   \\ 0 &  1/24 \end{array}\right];  
\bar S_{22}=\left[\begin{array}{cc}1 &  0   \\ 0 &  1 \end{array}\right];$  $\alpha_r=24.$ 
 $ 
U_{2}=  \left[\begin{array}{cc}1&0\\-1&2\end{array}\right];
$
$
L_3=  \left[\begin{array}{cc}0&0\\0&1\end{array}\right]; 
\\
 $
$L=  \left[\begin{array}{cccc}24 &  0 &  0 &  0 \\ 0 &  2 &  0 &  0 \\ 0 &   0 &1 &0 \\ 0 & 1 & 0 & 8 \end{array}\right]; 
S=  \ \left[\begin{array}{cccc}0 &  0 & \frac 1 {192} & 0 \\ \frac 1 2 & 0 &  0 &  0 \\ 0 &  0 &   0 &   0 \\ 0 & \ \frac 1 6 & 0 &  0 \end{array}\right];  
U=  \left[\begin{array}{cccc}2 & 0 & 1 &  0 \\ 0 & 8 & -1 & 2 \\ 0 & 0 & 24 & 0 \\ 0 & 0 & 0 &  1 \end{array}\right];$
$
\widehat S=\left[\begin{array}{cccc}0 &  0 & \frac 1 {4608} & 0 \\
\frac 1 {48} & 0 &  0 &  0 \\ 
0 &  0 &   0 & \frac 1 {24} \\
 0 & \frac  1 {384} & 0 &  0 \end{array}\right];
 $
$ M=\left[\begin{array}{cccc}0 & 24 & 0 & 0 \\ 0 & -24 & 0 & 48 \\ 192 & 0 &  0 &  0  \\ 0 &  0 & 24 & 0 \end{array}\right]; $ %
  $W=  \left[\begin{array}{cccc}-96 & 24 & 0 & 0 \\ 24 & 0 & -6 & 48 \\ 192 & 0 &  0 &   0  \\ 0 &   0 & 24 & 0 \end{array}\right].
  $

  \section{Conclusion}
 
This paper proposed a dichotomous recursive $LSU$ factorization algorithm.  
The algorithm has the complexity of matrix multiplication, and just like the LEU algorithm \cite{3}, it proves efficient for sparse matrices.
 
%

We can compare three algorithms LU, LEU, LSU. 
 
In algorithms LEU and LSU, the number of operations on the coefficients is of the same order as in the matrix multiplication algorithm they use (see Theorem 3 and \cite{3} Theorem 2), while algorithm LU has cubic complexity.

Consider  the case when the matrix coefficients are 64-bit real numbers. There are three possibilities. 

    (1) The matrix is well-conditioned and the accumulation of errors is insignificant.  In this scenario, the LEU algorithm will be the most optimal choice. It has the fewest number of operations when utilizing fast matrix multiplication algorithms.

    (2). When the matrix is not well-conditioned and the accumulation of errors is significant, the LU algorithm is the best choice, as it yields the smallest error. This is attributed to the fact that the pivot element is searched for among all coefficients of the matrix. Algorithm LEU is faster than LU; however, it accumulates more error.
    
    (3). When the matrix is ill-conditioned and no strategy provides an acceptable solution accuracy, both the LU and LEU algorithms are ineffective. In this situation, the LSU algorithm can be employed.
    
For matrices over integers or over polynomials, the factorization can only be performed using the LSU algorithm.




  \end{document}